\documentclass[%
reprint,
superscriptaddress,
amsmath,amssymb,
aps,
prl,
]{revtex4-2}
\usepackage{graphicx}
\usepackage{dcolumn}
\usepackage{physics} 
\usepackage{bm}
\usepackage[usenames,dvipsnames]{xcolor}

\definecolor{darkblue}{rgb}{0, 0, 0.8} 
\definecolor{hred}{HTML}{E43E4C} %
\definecolor{ptpurple}{HTML}{AA3377}
\definecolor{ptgreen}{HTML}{228833}

\usepackage{hyperref}
\hypersetup{colorlinks=true, breaklinks=true, linkcolor=ptpurple, citecolor=ptpurple, urlcolor=ptpurple, pdftitle={Dipole DSL}} 

\newcommand{\mb}{\mathbf}

\begin{document}

\title{ Dirac spin liquid in quantum dipole arrays}

\newcommand{\HarvardPhysicsAddress}{Department of Physics, Harvard University, Cambridge, Massachusetts 02138, USA}
\newcommand{\BerkeleyPhysicsAddress}{Department of Physics, University of California, Berkeley, CA 94720, USA}
\newcommand{\LBNLAddress}{Material Science Division, Lawrence Berkeley National Laboratory, Berkeley, CA 94720, USA}
\newcommand{\TUMAddress}{Technical University of Munich, TUM School of Natural Sciences,
	Physics Department, James-Franck-Str. 1, 85748 Garching, Germany}
\newcommand{\MunichAddress}{Munich Center for Quantum Science and Technology (MCQST), Schellingstr. 4, 80799 M\"unchen, Germany}
\newcommand{\CMUAddress}{Department of Physics, Carnegie Mellon University, Pittsburgh, PA 15213, USA}

\author{Marcus Bintz}
\affiliation{\HarvardPhysicsAddress}

\author{Vincent S. Liu}
\affiliation{\HarvardPhysicsAddress}

\author{Johannes Hauschild}
\affiliation{\TUMAddress}
\affiliation{\MunichAddress}

\author{Ahmed Khalifa}
\affiliation{\CMUAddress}

\author{Shubhayu Chatterjee}
\affiliation{\CMUAddress}

\author{Michael P. Zaletel}
\affiliation{\BerkeleyPhysicsAddress}
\affiliation{\LBNLAddress}

\author{Norman Y. Yao}
\affiliation{\HarvardPhysicsAddress}

\date{\today}

\begin{abstract}
	We predict that the gapless $U(1)$ Dirac spin liquid naturally emerges in a two-dimensional array of quantum dipoles.
	In particular, we demonstrate that the dipolar XY model---realized in both Rydberg atom arrays and ultracold polar molecules---hosts  a quantum spin liquid ground state on the kagome lattice.
	Large-scale density matrix renormalization group calculations indicate that this spin liquid
	exhibits signatures of gapless, linearly-dispersing spinons, consistent with the $U(1)$ Dirac spin liquid. 
	We identify a route to adiabatic preparation via staggered on-site fields and demonstrate that this approach can prepare cold spin liquids within experimentally realistic time-scales. 
	Finally, we propose a number of novel signatures of the Dirac spin liquid  tailored to near-term quantum simulators, including termination-dependent edge modes and the Friedel response to a local perturbation. 
\end{abstract}

\maketitle


Exotic phases of matter can emerge in deceptively simple quantum systems. 
For example, material electrons subject to a strong magnetic field and Coulomb interactions can fractionalize, forming 
quantum Hall liquids whose quasiparticle excitations carry a fraction of the original electron's quantum numbers~\cite{tsuiTwoDimensionalMagnetotransportExtreme1982, laughlinNobelLectureFractional1999}.
When such fractionalization occurs in an insulating spin system, the resulting phase of matter is known as a quantum spin liquid~\cite{savaryQuantumSpinLiquids2017}.
Although the theoretical viability of such phases is  well-established~\cite{readLargeNExpansionFrustrated1991, kitaevAnyonsExactlySolved2006}, their definitive identification and characterization remain a perennial challenge for both quantum materials and quantum simulators~\cite{knolleFieldGuideSpin2019,clarkQuantumSpinLiquids2021,altmanQuantumSimulatorsArchitectures2021}. 
On the latter front,  recent  works have explored several \emph{gapped} quantum spin liquids and their topological orders.
For example, a $\mathbb{Z}_2$ spin liquid may naturally arise from Rydberg  blockade interactions~\cite{verresenPredictionToricCode2021, semeghiniProbingTopologicalSpin2021}, while  digitally-operating quantum devices have probed wavefunctions with non-Abelian $D_4$ topological order~\cite{iqbalNonAbelianTopologicalOrder2024}. 

The spin liquids that emerge in conventional quantum materials are of a rather different sort.
In particular, studies of geometrically-frustrated antiferromagnets (e.g. on triangular, kagome, or pyrochlore lattices) often find indications of \textit{gapless} quantum spin liquids~\cite{shermanSpectralFunctionHeisenberg2023,drescherDynamicalSignaturesSymmetrybroken2023,huDiracSpinLiquid2019,xuRealizationDiracQuantum2023,wietekQuantumElectrodynamicsDimensions2024,heSignaturesDiracCones2017, liaoGaplessSpinLiquidGround2017,zhuEntanglementSignaturesEmergent2018,jiangCompetingSpinLiquid2019,zengSpectralEvidenceDirac2024, smithCaseQuantumSpin2022,smithQuantumSpinIce2023,poreeFractionalMatterCoupled2023,gaoEmergentPhotonsFractionalized2024,huangExtendedCoulombLiquid2020,chernPseudofermionFunctionalRenormalization2024}.
In two-dimensional systems, the exemplar is the  $U(1)$ Dirac spin liquid (DSL), whose low-energy excitations can be understood as massless, linearly-dispersing fermionic spinons coupled to an emergent $U(1)$ gauge field~\cite{ioffeGaplessFermionsGauge1989,marstonLargeLimitHubbardHeisenberg1989,hastingsDiracStructureRVB2000,wenQuantumOrdersSymmetric2002,ranProjectedWaveFunctionStudySpin2007}.
Its governing laws are those of a (2+1)-dimensional version of quantum electrodynamics  (QED$_3$), which represents a potentially stable, quantum critical phase of matter~\cite{appelquistHightemperatureYangMillsTheories1981,borokhovTopologicalDisorderOperators2002,hermeleStabilitySpinLiquids2004,hermelePropertiesAlgebraicSpin2008,songUnifyingDescriptionCompeting2019,albayrakBootstrappingConformalQED2022,heConformalBootstrapBounds2022,yeTopologicalCharacterizationLiebSchultzMattis2022,budarajuPiercingDiracSpin2023,karthikScalingDimensionFlux2024}.
%

\begin{figure}[ht!]
	\includegraphics{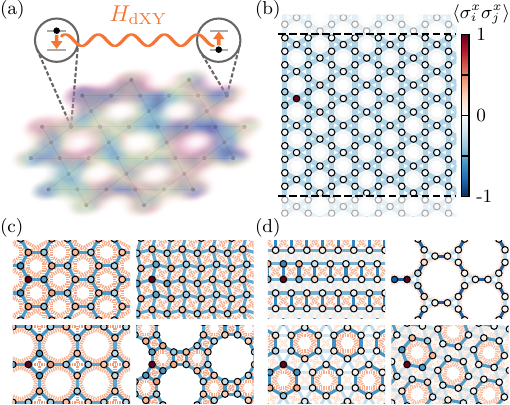}
	\caption{(a) Schematic depiction of a spin liquid emerging from the dipolar XY Hamiltonian in a two-dimensional array of easy-plane quantum dipoles.
		(b) Real-space spin-spin correlations of the ground state of $H_{\rm{dXY}}$ on the kagome lattice, obtained via  iDMRG on the YC12 cylinder. 
		Bonds between nearest- and next-nearest neighbors are colored according to  $\langle \sigma^x_i \sigma^x_j\rangle$. 
		Circles depict $\langle \sigma^x_i \sigma^x_0 \rangle$ correlations with a fixed site (dark red circle). 
		(c) Depicts spin-spin correlations on lattices where the ground state exhibits collinear ordering. 
		Dashed red bonds mark $\langle \sigma_i^x \sigma^x_j\rangle >0$. (d) Depicts 
		correlations on lattices where the ground state exhibits trivial local-singlets.
	}
	\label{fig:intro}
\end{figure}

%
In this work, we predict a new route to realize the gapless $U(1)$ Dirac spin liquid in synthetic quantum matter experiments.
Our proposal centers on a generic Hamiltonian that describes the interactions between coplanar quantum dipoles (Fig.~\ref{fig:intro}a). 
In particular, we consider effective spin-1/2 degrees of freedom, possessing a large transition dipole moment  but no permanent moment.
For a two-dimensional array of such objects, where the spin is quantized out of plane (i.e.~owing to a magnetic field), resonant dipole-dipole interactions yield the dipolar XY (dXY) Hamiltonian,
\begin{equation}\label{eq:HdXY}
H_{\rm{dXY}} =  \frac{J}{2} \sum_{i<j} \frac{\sigma^x_i \sigma^x_j + \sigma^y_i \sigma^y_j}{r_{ij}^3}.
\end{equation}
Here, $J$ is the interaction strength, $\vec{\sigma}$ are Pauli matrices, and $r_{ij}$ is the distance between spins $i$ and $j$.
This model is naturally realized in a wide variety of quantum simulators~\cite{hazzardQuantumCorrelationsEntanglement2014}, including ultracold polar molecules~\cite{baoDipolarSpinexchangeEntanglement2022, christakisProbingSiteresolvedCorrelations2023},  strongly-driven trapped ions~\cite{jurcevicQuasiparticleEngineeringEntanglement2014,richermeNonlocalPropagationCorrelations2014}, and Rydberg atom arrays~\cite{chenContinuousSymmetryBreaking2023}.

Our main results are threefold.
First, we conduct a large-scale infinite density matrix renormalization group (iDMRG)~\cite{whiteDensityMatrixFormulation1992,ostlundThermodynamicLimitDensity1995,mccullochInfiniteSizeDensity2008,schollwockDensitymatrixRenormalizationGroup2011,hauschildEfficientNumericalSimulations2018} investigation of $H_{\rm dXY}$ with antiferromagnetic interactions  on an extensive set of two-dimensional lattices (Fig.~\ref{fig:intro}).  
The majority exhibit either symmetry breaking or trivial paramagnetic ground states. 
However, on the kagome lattice, we observe a symmetric, highly-entangled  spin liquid.
The appearance of Dirac cones upon flux insertion suggests that this state is the $U(1)$ DSL.
Second, we identify a path to adiabatically prepare the DSL by applying a staggered on-site field, which effectively controls the mass of the Dirac spinons.
Dynamical simulations of our preparation protocol yield low-energy, liquid-like states within relatively short time-scales, $\tau \sim 10/J$. 
Finally, we propose and analyze a number of novel probes that can experimentally distinguish the $U(1)$ DSL from competing orders in  near-term quantum simulators: (i) negative spin susceptibilities, (ii) termination-dependent spinon edge modes  and (iii) the Friedel response to a local perturbation. 

\emph{Dipolar XY antiferromagnets.}---We compute the ground state of $H_{\rm{dXY}}$  on each of the eleven Archimedean tilings~\cite{grunbaumTilingsPatterns2016,tilingnote} using iDMRG on infinitely long cylinders with  circumference $W$~\cite{stoudenmireStudyingTwoDimensionalSystems2012,cincioCharacterizingTopologicalOrder2013,zaletelTopologicalCharacterizationFractional2013,szaszChiralSpinLiquid2020}.
We study states at half-filling of the conserved $U(1)$ charge, $M^z \equiv \sum_i \sigma^z_i = 0$, and include long-range couplings up to a maximum distance $R_{\rm{max}}\lesssim W/2$~\cite{supp}.
We find two main phases, and two exceptions.
One common state is collinear $U(1)$ symmetry breaking order,  evinced by long-range $\langle \sigma^x_i \sigma^x_j \rangle$ correlations with a clear Ne\'el sign structure~\cite{supp}.
This occurs when frustration is weak: on the square, hexagonal, truncated square, snub square, and truncated trihexagonal lattices [Fig.~\ref{fig:intro}(c)].
A second possibility arises for more frustrated systems but with an even number of spins per unit cell: the elongated triangular, truncated hexagonal,  rhombitrihexagonal,  and snub trihexagonal lattices [Fig.~\ref{fig:intro}(d)].
On these geometries, neighboring spins form local two- or six-spin singlet states, and the full many-body wavefunction is a trivial paramagnet (i.e.~to good approximation, a tensor product of the local singlets)~\cite{supp}.
The exceptional geometries are, perhaps unsurprisingly, the triangular and kagome lattices.
The former is somewhat sensitive to our numerical approximations and several phases closely compete in energy~\cite{toappear}.
%

\begin{figure}
	\includegraphics{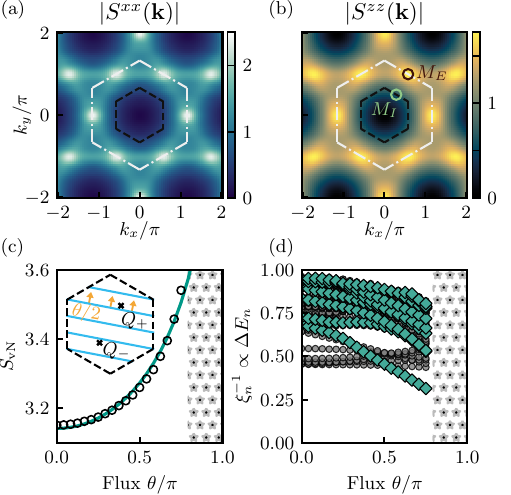}
	\caption{(a) The spin structure factor, $S^{xx}(\mb{k})$, exhibits diffuse weight near the perimeter of the extended Brillouin zone  (dash-dot gray hexagon), with some excess at the $M_E$ points. The inner Brillouin zone is shown as a black dashed hexagon. (b) Analogous structure factor for $S^{zz}(\mb{k})$. (c)  Adiabatically inserting an external flux $\theta$ shifts the spinon momentum bands towards the Dirac point (inset). The entanglement entropy diverges as $\theta$ approaches $\pi$ (YC8-2 cylinder, $d=7168$). Gray stars mark the region where adiabaticity is lost, and the teal line is a fit to Eq.~\ref{eq:SDirac}.  (d) The transfer matrix spectrum exhibits linearly dispersing modes. Teal points indicate eigenvalues with wavevectors corresponding to internode scattering. 
	}
	\label{fig:kagome} 
\end{figure}

\emph{Kagome dipolar XY spin liquid}.---On the kagome lattice, we find a robust, highly-entangled liquid.
We characterize this state on cylinders up to width $W=12$ (i.e.~the so-called YC12 geometry) and $R_{\rm max}\approx 3.7$, obtaining  good convergence (truncation error $\sim 2\times 10^{-5}$) at bond dimension  $d=10240$~\cite{supp}. 
The real-space $\langle \sigma^x_i \sigma^x_j \rangle$ correlations are depicted in Fig.~\ref{fig:intro}(b): their spatial uniformity and rapid decay indicate the absence of both spatial- and $U(1)$-symmetry-breaking order~\footnote{For distances greater than $W$, the decay is exponential.}.
This carries through to momentum space, where the equal-time spin structure factor,
$
S^{xx}(\mb{k}) =  \frac{1}{N}\sum_{i,j} e^{i \mb{k}\cdot(\mb{r}_i-\mb{r}_j)}  \langle \sigma^x_i \sigma^x_j \rangle,
$
exhibits diffuse weight around the extended Brillouin zone perimeter with some excess at the edge-centered $M_E$ points [Fig.~\ref{fig:kagome}(a)].
Interestingly, the $\sigma^z$-basis correlations show similar structure [Fig.~\ref{fig:kagome}(b)], even though the microscopic interactions are purely XY.  
For other observables (such as bond-bond correlations) and smaller cylinder circumferences, we find analogous results: all symmetries of $H_{\rm{dXY}}$ are unbroken~\cite{supp}.
Unlike the trivial paramagnets seen on other lattices, the Lieb-Schulz-Matthis-Oshikawa-Hastings theorems imply that our observed liquid must be non-trivial owing to the  three-site unit cell of the kagome lattice~\cite{liebTwoSolubleModels1961,oshikawaCommensurabilityExcitationGap2000,hastingsLiebSchultzMattisHigherDimensions2004}. 
The nature of this non-trivial liquid is perhaps hinted at by the relatively strong correlations at the $M_E$ points~\cite{zhuIdentifyingSpinonExcitations2019,zhangVariationalStudyGround2020}. 
Indeed, within the theory of the $U(1)$ DSL, such correlations in both $\sigma^x$ and $\sigma^z$ are natural, corresponding to spin-triplet monopole fluctuations of QED$_3$~\cite{hermelePropertiesAlgebraicSpin2008,songUnifyingDescriptionCompeting2019}.

\begin{figure}
	\includegraphics{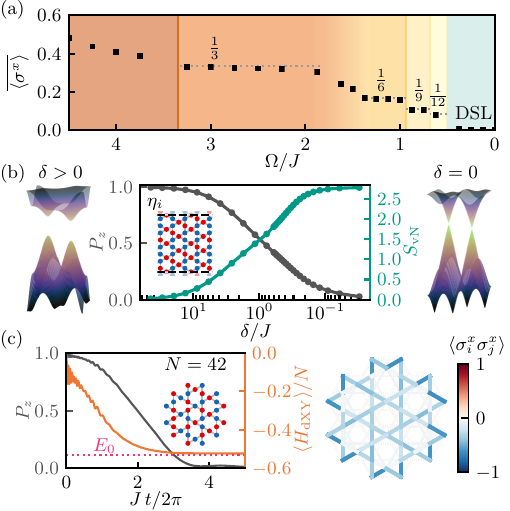}
	\caption{\label{fig:glider} (a) Depicts the phase diagram of $H_{\rm{dXY}} - \Omega \sum_i \sigma^x_i$ on the YC8 cylinder. A series of different phases (background color) are observed as quasi-plateaus in the average $\sigma^x$ magnetization. 
		(b) Depicts the  entanglement entropy and the staggered $\sigma^z$-polarization, $P_z$, as a function of $\delta$ for $H_{\rm{dXY}}+\delta H'$ on the YC8 cylinder.  
		There is a direct crossover from the critical $U(1)$ DSL at $\delta=0$ to the trivial paramagnet at large $\delta$.
		(Inset) The staggering pattern used on the YC8 cylinder. Red (blue) circles indicate $\eta_i=+1$ ($\eta_i=-1$); gray lines connect neighboring sites with opposite $\eta$, which have enhanced correlations at intermediate $\delta$.
		Left, right: illustration of the Dirac spinon band structure, which develops a mass gap when $\delta \ne 0$.
		(c) Depicts a TDVP simulation of the adiabatic preparation protocol:  $H_{\rm{dXY}} + \delta(t) H'$ evolution of an initial $\sigma^z$ product state (inset) using an exponential ramp $\delta(t)=\delta(0) e^{-t/\tau}$ with $\delta(0)=20\,J$ and $\tau = 0.6\times 2\pi/J$.  $P_z$ exhibits a smooth decay to zero, while the interaction energy approaches the DSL ground state energy, $E_0$ (dotted magenta line). 
		The final $\langle \sigma^x_i \sigma^x_j\rangle$ correlations are shown at right.
	}
\end{figure}

To further probe this $U(1)$ DSL hypothesis, we search for signatures of gapless spinons. 
In the kagome DSL, the spinon Dirac points are at $Q_\pm = \pm M_I/2$, and physical spin excitations (i.e.~two-spinon scattering processes) are gapless at the $\Gamma$ and $M_I$ points of the inner Brillouin zone [Fig~\ref{fig:kagome}(c), inset].
However, on a finite-width cylinder, the allowed spinon momentum bands  generically avoid the Dirac points, and all excitations  develop a $1/W$ gap~\cite{heSignaturesDiracCones2017, ferrariGaplessSpinLiquids2021}.
This can be overcome by simulated flux insertion: starting with a well-converged state on the YC8-2 cylinder, we slowly modify all  couplings by $\sigma_i^+\sigma_j^- \to e^{i\theta_{ij}}\sigma_i^+ \sigma_j^-$   and follow the changing ground state~\cite{heSignaturesDiracCones2017,zhuEntanglementSignaturesEmergent2018,huDiracSpinLiquid2019}.
Here,  $\theta_{ij}= (\mb{r}_{ij}\cdot \mb{r}_{\rm mod}/r_{\rm mod}^2) \,\theta  $, with $\mb{r}_{\rm{mod}}$ the periodic vector around the cylinder. 
This  flux insertion gradually shifts the spinon momenta by $\theta/2$, forcing them towards the Dirac points at $\theta =\pi$; we reach $\sim 3\pi/4$ before losing adiabatic continuity.
We track two key quantities as the flux is inserted.
The first is the  half-system von Neumann entanglement entropy, $S_{\rm vN}$,  which appears to diverge [Fig.~\ref{fig:kagome}(c)] consistent with the response expected from two Dirac cones,
\begin{equation}\label{eq:SDirac}
S_{\rm vN}(\theta) = A - B \sum_{i=0,1} \log \abs{2 \sin \frac{\theta-(-1)^i\pi}{2}},
\end{equation}
where $A, B$ are non-universal constants~\cite{zhuEntanglementSignaturesEmergent2018}. 
Second, we compute the iDMRG transfer matrix eigenvalues $\tau_n$: their magnitudes correspond to correlation lengths, $\xi_n = - 1/\log \abs{\tau_n}$, which are inversely proportional to excitation energies in critical systems with dynamical exponent $z=1$~\cite{zaunerTransferMatricesExcitations2015, ramsPreciseExtrapolationCorrelation2018, vanheckeScalingHypothesisMatrix2019, eberharterExtractingSpeedLight2023}.
The phases of $\tau_n$ correspond to wavevectors for these excitations.
As $\theta$ approaches $\pi$, we find that the eigenvalues with wavevectors near $M_I$ trend linearly downwards [Fig.~\ref{fig:kagome}(d)], suggestive of the linearly-dispersing gapless excitations characteristic of the DSL.

A few remarks are in order.
First, our iDMRG results are reminiscent of the quantum spin liquid seen in the paradigmatic nearest-neighbor kagome Heisenberg model, $H_{\rm{nn}} =  (J/2) \sum_{\langle i, j \rangle} \vec{\sigma}_i\cdot\vec{\sigma}_j$~\cite{elserNuclearAntiferromagnetismRegistered1989,misguichTWODIMENSIONALQUANTUMANTIFERROMAGNETS2013,heSignaturesDiracCones2017,zhuEntanglementSignaturesEmergent2018}.
Indeed, we find that the  ground state of $H_{\rm dXY}$ can be adiabatically connected to that of $H_{\rm{nn}}$, with no sign of any intervening phase transition~\cite{supp}.
Our results are also consistent with the observation of quantum spin liquids in the dipolar Heisenberg model~\cite{yaoQuantumDipolarSpin2018,zouFrustratedMagnetismDipolar2017,kelesAbsenceLongRangeOrder2018} and in short-range kagome XXZ models~\cite{lauchliQuantumSimulationsMade2015,heDistinctSpinLiquids2015,zhuChiralCriticalSpin2015,huVariationalMonteCarlo2015,changlaniMacroscopicallyDegenerateExactly2018}. 
Relatedly, we note that although the symmetries of $H_{\rm{dXY}}$ are reduced relative to $H_{\rm{nn}}$, they are still sufficient to forbid relevant perturbations to QED$_3$~\cite{supp, doddsQuantumSpinLiquids2013}.
%

\begin{figure*}
	\includegraphics{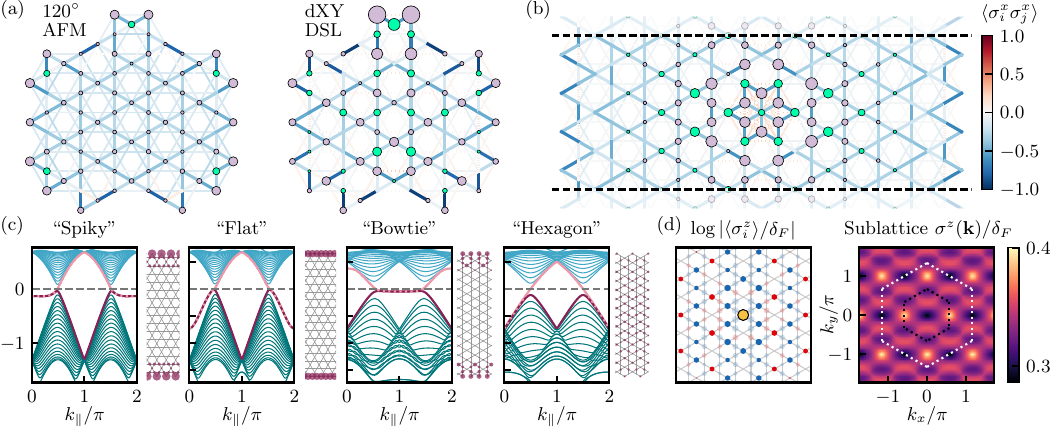}
	\caption{
		(a) DMRG ground state on an $N=83$ cluster for a short-range XY Hamiltonian~\cite{supp} in the 120$^\circ$ order phase (left) and for the $H_{\rm{dXY}}$ DSL (right). Bonds indicate $\langle \sigma^x_i \sigma^x_j\rangle$, and circles have size proportional to $\langle \sigma^z _i \rangle$ (lavender for $\langle \sigma^z_i \rangle >0$, green for $\langle \sigma^z_i \rangle<0$).  In the 120$^\circ$ state, the bulk spins uniformly cant up, while the $U(1)$ DSL exhibits strong oscillations. 
		(b) Similar oscillations can be induced by placing an interstitial spin at the center of a kagome hexagon.  (Additionally, this spin binds its twelve closest neighbors into a small collinear antiferromagnet).   (c)  Depicts edge physics of free Dirac spinons~\cite{supp} for different boundary terminations.  Left panels show the spectrum on an infinitely long strip with finite width.   Dark teal (light blue) lines indicate states below (above) half-filling. The two modes closest to half-filling are highlighted with dark purple and light pink lines. With the spiky, flat, and bowtie terminations, these zero-energy modes correspond to exponentially-localized edge states; the local density $|\psi_{i}|^2$ of the purple mode is shown on the right panels.  By contrast, the special ``hexagon'' boundary hosts no edge states: the center two modes are instead uniformly delocalized over the strip. (d) Left panel depicts free-spinon Friedel oscillations in response to a local impurity potential $\delta_F \sigma^z_0$. Red circles denote $\langle \sigma^z_i\rangle >0$, while blue circles denote $\langle \sigma^z_i \rangle < 0$, with the size scaling as $\log |\langle \sigma^z_i \rangle |$. Semi-transparent markers are placed for sites on sublattices different than $\mb{r}_0$. Right panel: Fourier transform of $\langle \sigma^z_i\rangle$, restricting to sites on the same sublattice as $\mb{r}_0$. The dominant weight is at $M_I$.
	}
	\label{fig:static}
\end{figure*}

\emph{Adiabatic preparation.}---We now turn to the question of how to prepare  the Dirac spin liquid. 
Contemporary quantum simulators are well-isolated systems, so correlated states must generally be prepared in a dynamical fashion~\cite{albashAdiabaticQuantumComputation2018}.
The criticality of the $U(1)$ DSL suggests a conceptually simple adiabatic strategy: apply a relevant perturbation~\cite{chandranKibbleZurekProblemUniversality2012}.
In the ideal case, the ground state can be smoothly followed between the low- and high-strength limits of the perturbing field.
As the applied field, $\delta$, is decreased towards the critical phase (at $\delta=0$), the correlation length will diverge as $\xi\sim \delta^{-\nu}$ until either adiabaticity is lost or $\xi$ exceeds the linear system size.
In fact, adiabaticity can be maintained for all times if $\delta(t)$ decreases asymptotically as $t^{-p}$ with $p< 1/(\nu z)$~\cite{chandranKibbleZurekProblemUniversality2012}.

The problem is then to identify a relevant perturbation that is experimentally feasible and does not  lead to intervening phases at intermediate field strengths. 
Perhaps the most straightforward possibility is to apply a uniform  field, $\Omega  \sum \sigma^x_i$, i.e.~a resonant, global driving field; this corresponds to a conserved $SU(4)$ current of QED$_3$ with scaling dimension $\Delta_J = 2$. 
However, in iDMRG we find a long cascade of phase transitions between the $U(1)$ DSL and the large-field paramagnet [Fig.~\ref{fig:glider}(a)], prohibiting this as a route for adiabatic preparation.
Instead, we propose a spatially modulated transverse field,
$H' = -\sum_i \eta_i \sigma^z_i$,
where $\eta_i = \pm 1$ is a binary staggering pattern~\cite{jakschColdBosonicAtoms1998,priyadarsheeQuantumPhaseTransitions2006,motrukPhaseTransitionsAdiabatic2017}.
We note that such a perturbation was recently implemented in a Rydberg-based  dipolar XY experiment using light-shifts from local addressing~\cite{chenContinuousSymmetryBreaking2023, bornetEnhancingManybodyDipolar2024}. 
Theoretically, it corresponds to a mass term for the Dirac fermions: a relevant operator with scaling dimension $\Delta_N \approx 1.4$, which is ordinarily forbidden by spatial and spin-flip symmetries that $H'$ explicitly breaks~\cite{hermelePropertiesAlgebraicSpin2008,supp}.
This mass gaps out the Dirac spinons, in turn triggering a confinement transition of the compact $U(1)$ gauge field into a trivial gapped paramagnetic phase~\cite{polyakovQuarkConfinementTopology1977,polyakov1987gauge}.
With the proper choice of $\eta_i$, we find that the ground state phase diagram of $H_{\rm{dXY}} + \delta H'$ indeed exhibits a smooth crossover from the DSL to the high-field paramagnet, as shown in Fig.~\ref{fig:glider}(b)~\footnote{This is a finite-$W$ smoothing of the singular behavior expected as $\delta\to 0$ in the thermodynamic limit.}. 
Indeed, the entanglement entropy and the staggered $\sigma^z$-polarization, $P_z = \sum_i \eta_i \langle \sigma^z_i \rangle$, smoothly interpolate between the $\delta=0$ [$U(1)$ DSL] and the $\delta\to\infty$ (trivial paramagnet) limits~\cite{supp}. 
In principle, then, this is a promising route for adiabatic preparation.

To test this, we utilize  a time-dependent variational principle calculation~\cite{haegemanTimeDependentVariationalPrinciple2011,haegemanUnifyingTimeEvolution2016} to directly simulate preparation  dynamics on an open boundary kagome cluster with $N=42$ spins [Fig.~\ref{fig:glider}(c)]. 
Beginning with the product state shown in Fig.~\ref{fig:glider}(c, inset),  we evolve under the Hamiltonian, $H(t) = H_{\rm{dXY}}+\delta(t)H_{\eta\rm{Z}}$.
Starting with $\delta(0) = 20\,J$, we  exponentially ramp down the field, $\delta(t) = \delta(0) e^{-t/\tau}$ with $\tau=0.6\times 2\pi/J$ up to a final preparation time $T$. 
Even for short preparation time-scales, $T = 5 \times 2\pi /J$, we observe that  $P_Z$ smoothly reduces towards zero and the final spin-spin correlation functions are relatively uniform in the bulk [Fig.~\ref{fig:glider}(c)]. 
Moreover, by the end of the ramp, the energy of the state is within a few percent of the DSL ground state value ($0.984\, E_0$).

\emph{Signatures of the DSL in quantum simulators}---We finally come to a common challenge for the experimental detection of a spin liquid: how to  tell?
In principle, the presence of gapless, fractionalized spinon excitations will influence most equilibrium and dynamical observables; for instance, generic correlation functions should decay as power-laws in space and time~\cite{hermelePropertiesAlgebraicSpin2008}. 
In this sense, the $U(1)$ DSL may actually be somewhat easier to probe than a gapped quantum spin liquid.
However, there are two key challenges for small- and intermediate-size experiments: (i) limited length scales to cleanly resolve power-laws, and (ii) the presence of an edge.
In particular, the low-energy, spin-singlet modes of QED$_3$ with finite momentum  will generally condense into valence bond solids at the boundary~\cite{mcavityConformalFieldTheories1995,diehlTheoryBoundaryCritical1997,songUnifyingDescriptionCompeting2019,hermelePropertiesAlgebraicSpin2008}.
These edge correlations will decay into the liquid bulk as a power-law, $r^{-\Delta}$, which complicates the interpretation of many observables~\footnote{We note that $\Delta$ may be as low as $\Delta_\Phi\approx1.1$ for monopole excitations.}.

Nevertheless, there are positive signatures of the DSL that we expect can be probed even on existing quantum simulators (with $N \sim 100$). 
In fact, the spin structure factor, $S^{xx}(\mb{k})$, already provides some insight: the monopole weight at the $M_E$ point can be seen on moderately sized clusters, and its presence allows one to rule out competing valence bond solids with different structure factors~\cite{supp}. 
In what follows, we describe three additional DSL signatures. 
First, we note that the aforementioned $M_E$ signal could just as well indicate coplanar 120$^\circ$ magnetic order---a natural and oft-seen competing phase near the Dirac spin liquid~\cite{songUnifyingDescriptionCompeting2019}. 
One probe that cleanly cuts between the DSL and the 120$^\circ$ state is as follows:  measure $\langle \sigma^z_i\rangle$ in systems with an excess spin 1/2, i.e. on odd-$N$ clusters.
A coplanar magnet will uniformly cant its bulk spins upward, leading to $\langle \sigma^z_i \rangle \sim 1/N$; an example of this behavior for a short-ranged XY model is illustrated in  Fig.~\ref{fig:static}(a)~\cite{supp}.
By contrast, we observe that the local magnetization in the spin-doped DSL is highly oscillatory, with many spins exhibiting a large, negative response, $\langle \sigma^z_i \rangle <0$ [Fig.~\ref{fig:static}(b)]~\cite{gregorNonmagneticImpuritiesSpin2008, yangStrongRelevanceZinc2024}.
Relatedly, when an interstitial spin is placed at the center of a kagome hexagon, we again observe that the  excess charge produces large local oscillations in $\langle \sigma^z_i\rangle$ [Fig.~\ref{fig:static}(c)].

The other two signatures we propose to explore in near-term experiments are both related to the presence of fermionic spinons in the DSL. 
To start, in the standard, mean-field spinon theory of the $U(1)$ DSL~\cite{hastingsDiracStructureRVB2000,ranProjectedWaveFunctionStudySpin2007,supp}, we find that generic boundaries of the kagome lattice will host localized edge modes at the Fermi level; the  exception is a special ``hexagon'' termination  where the zero-energy excitations instead uniformly spread over the system  [Fig.~\ref{fig:static}(d)].
This is reminiscent of the termination-dependent edge modes seen in graphene~\cite{fujitaPeculiarLocalizedState1996,nakadaEdgeStateGraphene1996}, and is intimately related to the presence of Dirac cones~\cite{breyElectronicStatesGraphene2006,biswasMasslessFermionsHalfspace2022,supp}.
The ability of optical tweezer arrays to realize arbitrary geometries provides a natural testbed for this physics; for instance, by exploring which types of edges can coherently propagate a flipped spin~\cite{braunRealspaceDetectionManipulation2023}.
Finally, applying a \emph{local} static perturbation $\sigma_0^z$ should induce $M_I$-point Friedel oscillations from internode spinon scattering [Fig.~\ref{fig:static}(e)], i.e., a particle-hole excitation between the DSL's two different Dirac points separated by $M_I$.
These manifest in the single-body $\langle \sigma^z_i\rangle$ static response, which is  particularly apt for experiments with single-site resolution.

Looking forward, there are a number of intriguing direction to further investigate. 
First, it is important to understand the stability of the DSL to   positional disorder~\cite{thomsonQuantumElectrodynamicsDimensions2017,deyDestabilizationDiracSpin2020,seifertSpinPeierlsInstabilityDirac2023,ferrariSpinphononInteractionsKagome2024} and missing spins~\cite{olariu17NMRStudy2008,gregorNonmagneticImpuritiesSpin2008,poilblancImpuritydopedKagomeAntiferromagnet2010,yangStrongRelevanceZinc2024}, both of which are present in near-term simulators.
Second, the dipolar XY Hamiltonian has no free parameters, but can be tuned by modifying the lattice geometry;  possible applications of this geometric control could be to reduce edge effects or to study line defects in the DSL~\cite{vekicSmoothBoundaryConditions1993,katsuraSinesquareDeformationSolvable2012,billoDefectsConformalField2016,huSolvingConformalDefects2024}. 
Finally, we note that the spinon-edge-state and Friedel oscillation signatures are both mean-field spinon predictions---understanding possible modifications from the dynamical $U(1)$ gauge field remains an important open question~\cite{kolezhukTheoryQuantumImpurities2006,ranSpontaneousSpinOrdering2009,herzogFermionsBoundaryConformal2023}.

\emph{Acknowledgments.}---We gratefully acknowledge the insights of  A. Browaeys, G. Bornet, C. Chen, G. Emperauger, J. Kemp, T. Lahaye, K.~K. Ni, P. Scholl, R. Verresen, A. Vishwanath, and J. Wei.
This work was supported in part by the U.S. Department of Energy, Office of Science, National Quantum Information Science Research Centers, Quantum Systems Accelerator and by the Air Force Office of Scientific Research via the MURI program (FA9550-21-1-0069).
V.L.~acknowledges support from the NSF through the Center for Ultracold Atoms.
J.H.~was funded by the U.S. Department of Energy, Office of Science, Office of Basic Energy Sciences, Materials Sciences and Engineering Division under Contract No. DE-AC02-05- CH11231 through the Scientific Discovery through Advanced Computing (SciDAC) program (KC23DAC Topological and Correlated Matter via Tensor Networks and Quantum Monte Carlo).
M.Z.~was  supported primarily by the U.S. Department of Energy, Office of Science, Basic Energy Sciences, under Early Career Award No. DE-SC0022716.
N.Y.Y.~acknowledges support from a Simons Investigator award. 
\bibliography{kgm_paper, footnotes_and_other}

\newpage

\end{document}


\newcommand{\HarvardPhysicsAddress}{Department of Physics, Harvard University, Cambridge, Massachusetts 02138, USA}
\newcommand{\BerkeleyPhysicsAddress}{Department of Physics, University of California, Berkeley, CA 94720, USA}
\newcommand{\LBNLAddress}{Material Science Division, Lawrence Berkeley National Laboratory, Berkeley, CA 94720, USA}
\newcommand{\TUMAddress}{Technical University of Munich, TUM School of Natural Sciences,
	Physics Department, James-Franck-Str. 1, 85748 Garching, Germany}
\newcommand{\MunichAddress}{Munich Center for Quantum Science and Technology (MCQST), Schellingstr. 4, 80799 M\"unchen, Germany}
\newcommand{\CMUAddress}{Department of Physics, Carnegie Mellon University, Pittsburgh, PA 15213, USA}

\author{Marcus Bintz}
\affiliation{\HarvardPhysicsAddress}

\author{Vincent S. Liu}
\affiliation{\HarvardPhysicsAddress}

\author{Johannes Hauschild}
\affiliation{\TUMAddress}
\affiliation{\MunichAddress}

\author{Ahmed Khalifa}
\affiliation{\CMUAddress}

\author{Shubhayu Chatterjee}
\affiliation{\CMUAddress}

\author{Michael P. Zaletel}
\affiliation{\BerkeleyPhysicsAddress}
\affiliation{\LBNLAddress}

\author{Norman Y. Yao}
\affiliation{\HarvardPhysicsAddress}

\title{Supplemental Material for ``Dirac spin liquid in quantum dipole arrays''}
\date{\today}

\maketitle
\tableofcontents

\newpage
In this Supplemental Material, we provide additional methodological details, numerical results, and analytical arguments that support the conclusions in the main text.
In the first two sections, we discuss analytic aspects of the $U(1)$ DSL and its corresponding free-spinon mean-field theory.
The next two sections give additional numerical details and results for the $H_{\rm{dXY}}$ model on the kagome lattice.
The final sections discuss other lattice geometries and some aspects of the adiabatic preparation, and are mostly independent of the preceding discussion.

\section{The \texorpdfstring{$U(1)$}{U(1)} Dirac spin liquid: theoretical aspects}\label{app:dsl}

We first review some basic properties of the $U(1)$ DSL and establish some notation.
The $U(1)$ DSL is a fractionalized, quantum critical state of matter whose low-energy, long-wavelength excitations can be understood as massless Dirac fermions, $\psi$, coupled to a compact $U(1)$ gauge field, $A_\mu$. 
Their dynamics are governed by the Lagrangian of (2+1)-dimensional quantum electrodynamics (QED$_3$),
\begin{equation}
\mathcal{L}_{\mathrm{QED}} = \bar{\Psi}\left[-i\gamma^\mu(\partial_\mu+iA_\mu)\right] \Psi + \frac{1}{2e^2}\sum_\mu \left(\epsilon_{\mu\nu\lambda} \partial_\nu A_\lambda\right)^2.
\end{equation}
Here, we follow the conventions of Ref.~\cite{hermelePropertiesAlgebraicSpin2008}: $\Psi$ is a vector of $N_f$ two-component Dirac fermions, $\psi$; $\gamma^\nu=(\tau^3,\tau^2,-\tau^1)$ for $\nu=0,1,2$, where $\tau$ are Pauli matrices acting on the two-component Dirac space; $\bar{\Psi}=i\Psi^\dagger \tau^3$; and $e$ is the coupling constant.
For a spin-1/2 lattice system with two Dirac cones, $N_f=4$ and $\mathcal{L}_{\mathrm{QED}}$ possesses a global $SU(4)$ symmetry generated by $T^a =  \left\{\sigma^i, \mu^i, \sigma^i\mu^j\right\}$, where $\sigma$ acts on the spin degree of freedom, and $\mu$ on the valley indices.
The essential physical consequence of this $SU(4)$ symmetry is that microscopically very different operators can exhibit exactly the same scaling dimensions~\cite{hermeleStabilitySpinLiquids2004} (and, consequently, power-law decays).
Lastly, $\mathcal{L}_{QED}$ exhibits an emergent global $U(1)$ symmetry corresponding to the total magnetic flux through the plane; in a spin system, the $U(1)$ gauge field is compact, so this flux is only conserved modulo $2\pi$~\cite{hermeleStabilitySpinLiquids2004}.
Field operators that change this flux (i.e. when acting on the QED$_3$ vacuum) by units of $2\pi$ are referred to as monopoles~\cite{hermeleStabilitySpinLiquids2004,songUnifyingDescriptionCompeting2019}.

As an effective field theory, $\mathcal{L}_{\mathrm{QED}}$ should be supplemented with a series of all symmetry-allowed local terms.
If all such terms are irrelevant,  $\mathcal{L}_{\mathrm{QED}}$ will flow in the infrared (IR) to a strongly-coupled conformal field theory (CFT), and thus represents a stable critical phase (i.e. without fine tuning).
The most important classes of perturbations are fermion billinear (mass) terms, and monopole insertion operators.
In the kagome Heisenberg antiferromagnet, all fermion billinears and $2\pi$-monopole operators  are forbidden by the global symmetries of the microscopic model~\cite{hermelePropertiesAlgebraicSpin2008,songUnifyingDescriptionCompeting2019}.
The stability of the phase then depends on the scaling dimension of the $4\pi$-monopole operator: the most recent lattice monte carlo estimate for $N_f=4$ QED$_3$ is $\Delta_{4\pi}=3.7(3)$, which is irrelevant and so implies that the $U(1)$ DSL phase is possibly stable~\cite{karthikScalingDimensionFlux2024}.
We note there is currently some tension between lattice monte carlo and conformal bootstrap approaches~\cite{albayrakBootstrappingConformalQED2022,heConformalBootstrapBounds2022}.

\begin{figure*}
	\includegraphics{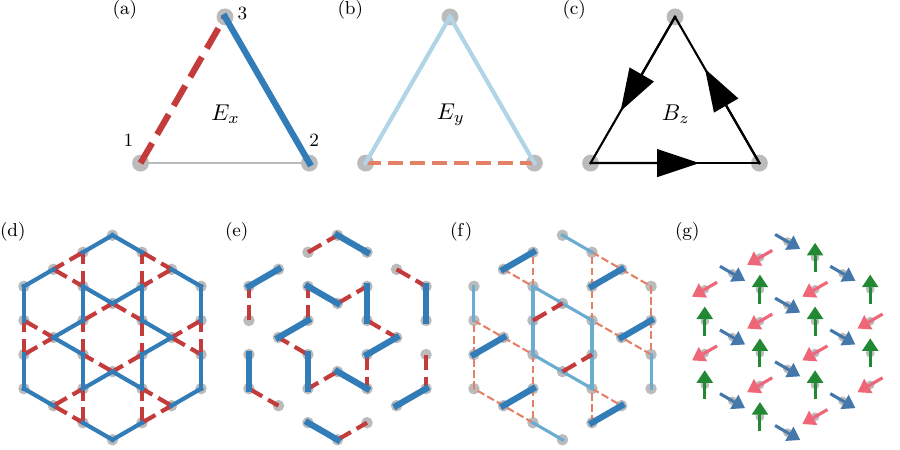}
	\caption{\label{fig:SI_ope} Illustration of leading microscopic operators that are symmetry-equivalent to the emergent $U(1)$ gauge fields (top row) and to the  $2\pi$-monopoles (bottom row)~\cite{hermelePropertiesAlgebraicSpin2008,thomsonQuantumElectrodynamicsDimensions2017,songUnifyingDescriptionCompeting2019, nambiarMonopoleJosephsonEffects2023}.  (a,b) The electric field corresponds to spatial anisotropy in spin-spin correlations $\langle \vec{\sigma}_i\cdot\vec{\sigma}_j\rangle$. (c) The leading magnetic field operator is the three-spin scalar chirality $\langle \vec{\sigma}_i\cdot(\vec{\sigma}_j\times\vec{\sigma}_k)\rangle$ about a fundamental kagome triangle. (d, e, f) The spin-singlet monopoles $\Phi_s$ carry finite momentum and correspond to different valence bond solid patterns.  (g) The spin-triplet monopole $\Phi_t$ induces coplanar $120^\circ$ magnetic order. This ``$q=0$ state'' breaks spatial reflection and rotation, but not translation.
	}
\end{figure*}

The emergent long-wavelength fields can be related to microscopic physical observables through symmetry arguments and the operator product expansion (OPE).
Precisely, if a physical operator $\mathcal{O}$ has the same symmetries as some emergent field $\phi$ with scaling dimension $\Delta$, then it should generally exhibit two-point correlations that decay spatially as $r^{-2\Delta}$.
If $\phi$ has nontrivial momentum then the decay will also oscillate at that wavevector.
The microscopic correspondences of several important kagome DSL fields have been worked out previously~\cite{hermelePropertiesAlgebraicSpin2008,thomsonQuantumElectrodynamicsDimensions2017,songUnifyingDescriptionCompeting2019, nambiarMonopoleJosephsonEffects2023}, and some are illustrated in Fig.~\ref{fig:SI_ope}.

\subsection{Stability of the DSL with dipolar XY interactions}
There are two key ways that $H_{\rm dXY}$ differs from $H_{\rm nnH}$ which could alter the DSL stability argument: the spin-rotation symmetry is reduced ($U(1)\rtimes  \mathbb{Z}_2 \subset SU(2)$), and the interactions are somewhat long-ranged.
The first point could enable a relevant operator that was previously symmetry-forbidden to become symmetry-allowed.
We argue this does not end up happening.
Besides the $U(1)\rtimes \mathbb{Z}_2$ rotation symmetry, $H_{\rm{dXY}}$ also has the antiunitary time reversal symmetry $\vec{\sigma}\to -\vec{\sigma}$, as well as all spatial symmetries of the kagome lattice.
As stated earlier, the potentially relevant operators are the fermion billinears and the $2\pi$ monopoles.
The understood symmetry transformations of these fields are summarized in Table 3 of Ref.~\cite{songUnifyingDescriptionCompeting2019}, from which one can immediately verify that all such operators transform nontrivially under at least one of the $H_{\rm{dXY}}$ symmetries.
As an alternative perspective, in the mean-field spinon framework (see Sec.~\ref{app:spinon}), one can explicitly determine all symmetry-allowed spinon billinears; this was done in Ref.~\cite{doddsQuantumSpinLiquids2013}, and in the time-reversal symmetric $(D=0)$ case of their Eq. (18), (19),  the DSL is seen to be mean-field stable.
The second point is also not expected to be an issue, because the decay of the  $r^{-3}$ interaction and of the DSL spin-spin correlations are sufficiently fast.
More explicitly, we can follow a general argument by Cardy~\cite{cardyScalingRenormalizationStatistical1996}: consider a short-range Hamiltonian $H_0$ which is critical, and perturb it with some long-range interaction $H_1 = \sum_{r,r'}V(r-r')\mathcal{O}(r)\mathcal{O}(r')$, where $V(r)\sim v_0 r^{-d-\sigma_V}$, with $d$ the spatial dimension.
We then determine whether this is a relevant perturbation to the short-range fixed point.
Define $q(R)\equiv \int \mathrm{d}^dr\; V(r) \mathcal{O}(R+r/2)\mathcal{O}(R-r/2)$, so that
\begin{equation}
\langle q(R_1) q(R_2) \sim \int \int V(r_1) V(r_2) \langle \mathcal{O}(R_1+r_1/2)\mathcal{O}(R_1-r_1/2) \mathcal{O}(R_2+r_1/2)\mathcal{O}(R_2-r_1/2)\rangle \dd^dr_1\dd^dr_2.
\end{equation}
The scaling of this expression with $R_{12}=R_1-R_2$ follows from the scaling behavior of each individual component: it is $R_{12}^{2d-2(d+\sigma_V)-4\Delta_\phi}$, where $\phi$ is the most relevant operator contributing to $\mathcal{O}$ correlations. 
The scaling dimension of the perturbing interaction is then $\Delta_q = 2 \Delta_\phi + \sigma$.
In our $H_{\rm{dXY}}$ setting, $\sigma_V=1$, and the most relevant operator related to $\sigma^+\sigma^-$  correlations is the spin-triplet $2\pi$ monopole $\Phi_t$~\cite{hermelePropertiesAlgebraicSpin2008,songUnifyingDescriptionCompeting2019}.
This operator is generally found to have dimension slightly greater than 1~\cite{heConformalBootstrapBounds2022,albayrakBootstrappingConformalQED2022}, for example, the monte carlo study Ref.~\cite{karthikNumericalDeterminationMonopole2019} gives $\Delta_\Phi = 1.26(8)$.
The $r^{-3}$ spin-spin interactions of $H_{\rm{dXY}}$ are thus expected to be renormalization group irrelevant ($\Delta_q \approx 3.5>3$), leaving the $U(1)$ DSL fixed point in principle stable. 
Two additional remarks: (i) The stability of the $U(1)$ DSL for short-range kagome Heisenberg interactions remains somewhat ambiguous~\cite{heConformalBootstrapBounds2022,albayrakBootstrappingConformalQED2022,karthikScalingDimensionFlux2024}. Our claim is that $H_{\rm{dXY}}$ does not introduce any new essential instabilities.  (ii) Generically, microscopic energetics could instead favor a spontaneous-symmetry-breaking phase, i.e. into one of the proximate, descendant orders of the $U(1)$ DSL~\cite{hermeleAlgebraicSpinLiquid2005,hermelePropertiesAlgebraicSpin2008,songUnifyingDescriptionCompeting2019}. 
Numerically, we were able to stabilize several such states by adjusting the strength of various long-range couplings (see Sec.~\ref{sec:SI_global}).
However, our DMRG results indicate  $H_{\rm{dXY}}$ instead lies in the critical, symmetry-unbroken phase.

\subsection{Mean-field spinon theory}\label{app:spinon}

\begin{figure}
	\includegraphics{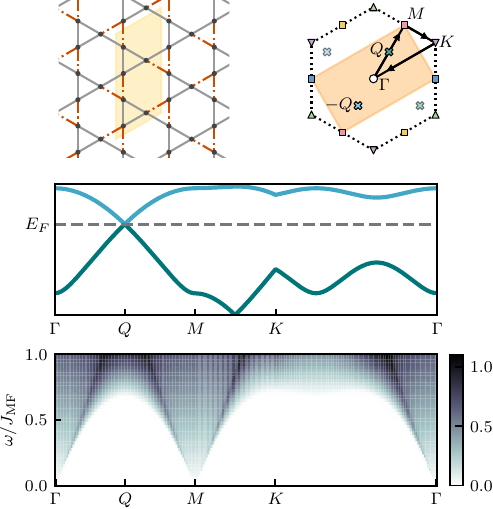}
	\caption{\label{fig:SI_spinonschematic}  Basic properties of the $(\pi,0)$ model for free Dirac spinons on the kagome lattice. (a) One choice of unit cell (light orange parallelogram) and gauge (red dash-dot lines). (b) The spinon BZ is half of the inner BZ, and contains two Dirac points, $\pm Q$. (c) Center two bands, $E_{\rm{MF}}(\mb{k})$, of the $(\pi,0)$ model, along the momentum cut shown with black arrows in (b). (d) The  dynamical spin structure factor, $\mathcal{S}_{{\mathfrak{a}\mathfrak{a}}}(\mb{k},\omega)$, is a  gauge-invariant convolution of two spinons, and so displays conical weight at the  $\Gamma$ and $M = 2 Q$ points. 
	}
\end{figure}

An alternative way to understand the $U(1)$ DSL is through the mean-field ``parton theory'' approach~\cite{savaryQuantumSpinLiquids2017}, in which one fractionalizes the local spin-1/2 operators into pairs of fermionic spinons
\begin{equation}
S^+_i= f_{i,\uparrow}^\dagger f_{i,\downarrow}, \quad S^-_i = f_{i,\downarrow}^\dagger  f_{i,\uparrow}, \quad \sigma^z_i = n_{i,\uparrow}-n_{i,\downarrow},
\end{equation}
where $f$ are fermionic ladder operators, and $n= f^\dagger f$.
If one imposes the single occupancy constraint, $n_{i,\uparrow}+n_{i,\downarrow}= 1$, then the spin and constrained fermion Hilbert spaces and operator algebras are isomorphic.
In the mean-field approximation, this local constraint is relaxed to be true only on average (i.e. the fermions are at half-filling), and different possible spin liquid states are then represented by gauge-inequivalent quadratic spinon Hamiltonians~\cite{wenQuantumFieldTheory2010}.
The one for the kagome DSL takes the pure-hopping form,
\begin{equation}
H_{\rm \sigma MF} = - t \sum_{\sigma = \uparrow,\downarrow} \sum_{\langle i j \rangle} s_{ij} f^\dagger_{\sigma i} f_{\sigma j},
\label{eq:HmfPiFlux}
\end{equation} 
where $t$ is a uniform hopping and $s_{ij} = \pm 1$ is understood as follows: writing $w_{ij} = e^{iA_{ij}}$,  $A_{ij}$ can be interpreted as (the spatial components of) an emergent $U(1)$ gauge field, whose mean-field configuration on the kagome lattice has  $0$ flux through elementary triangles, and $\pi$ flux through elementary hexagons.
One choice of gauge for this flux pattern is shown in Fig.\,~\ref{fig:SI_spinonschematic}(a); other gauges are related by local unitary transformations, $f_{j,\sigma}^\dagger \to e^{i\theta_j} f_{j,\sigma}^\dagger$.
On the infinite plane, $H_{\rm \sigma MF}$  is reduced to a six-band model, for which the two center bands  meet at a pair of Dirac points, $\pm{Q}=\pm M_3/2$ [Fig.~\ref{fig:SI_spinonschematic}(b,c)].
(Due to the $\pi$ fluxes, the spinon Brillouin zone (sBZ) is necessarily halved, reflecting the projective symmetry group of the fermions.)
We note that in the mean-field approximation, the up- and down-spinons do not interact, so it is often convenient to focus on a single species of spinon, i.e. work with the spinless Hamiltonian $H_{\rm MF} = -  \sum_{\langle i,j \rangle} t_{ij} f_{i}^\dagger f_{j}$

\subsection{Mass term for staggered \texorpdfstring{$\sigma^z$}{Z} field}

\begin{figure*}
	\includegraphics{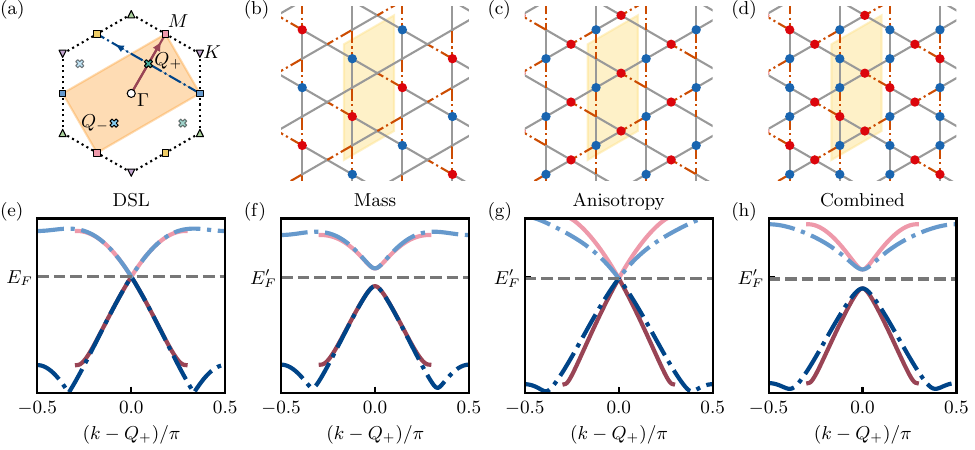}
	\caption{\label{fig:SI_glidermass} (a) Inner Brillouin zone (hexagon), magnetic spinon Brillouin zone (rectangle), high-symmetry points, and two orthogonal momentum cuts that pass through a spinon Dirac point $Q_+$.  (b-d) The $\sigma^z$  staggering pattern studied in the main text can be decomposed into a $q=M_I$ part and a translation-invariant part. The lattice bonds depict a compatible set of amplitudes for mean-field spinons (see Fig.~\ref{fig:SI_spinonschematic}). (e) Mean-field spinon band structure in the absence of the staggered field. Blue and red lines correspond to the two different cuts in (a).  Dark/light lines are the bands below/above the Fermi level. The Dirac dispersion is isotropic, as seen by the overlap of the red and blue lines. (f) The $q=M_I$ contribution opens a mass gap at the Dirac point. (g) The translation-invariant part introduces a velocity anisotropy. (h) Combined effect of the two terms.
	}
\end{figure*}

We now discuss how the staggered $\sigma^z$ field introduced in the main text can be understood from the continuum theory.
Here we write this perturbation as:
\begin{equation}
H_z = -\sum_i \eta_i \sigma_i^z,
\end{equation}
where $\eta_i=\pm 1$ is the binary staggering pattern, which we reproduce in Fig.~\ref{fig:SI_glidermass}(d).
(In the main text, this perturbation was denoted $H'$.)
We then split $H_z$ into a sum of two terms, $H_z = H_z^{(1)} + H_z^{(2)}$,  where the $\eta$ for $H_z^{(1)}, H_z^{(2)}$, are shown in Fig.~\ref{fig:SI_glidermass}(b,c).
The first term, $H_z^{(1)}$, is nonzero on a single sublattice and alternates in sign at $q=M_I$.
As pointed out in Ref.~\cite{hermelePropertiesAlgebraicSpin2008} (c.f. Fig. 4 in that work), this  exactly corresponds to the fermion mass billinear $(\mb{N}^2_A)^z = \Psi^\dagger \tau^3 \mu^2 \sigma^z\Psi$.
This also holds in the standard mean-field spinon description, where $\sigma^z_i = n_{i,\uparrow} - n_{i,\downarrow}$ corresponds to an onsite chemical potential of opposite sign for spin-up vs. spin-down spinons.
Representing $H_z^{(1)}$ in this way and adding it to the gapless mean-field Hamiltonian (Eq.~\ref{eq:HmfPiFlux}) directly opens a mass gap at the Dirac point [Fig.~\ref{fig:SI_glidermass}(f)].
The second, translation-invariant term is somewhat more subtle, but its effect can be determined through a similar mean-field analysis.
As shown in Fig.~\ref{fig:SI_glidermass}(g), this term does \textit{not} open a gap, but instead leads to an anisotropic dispersion relation, which can be seen by considering two orthogonal cuts through momentum space. 
The anisotropy is of opposite sign for up/down spinons (not shown), and the corresponding continuum operator is the kinetic term $K_a = -i \Psi^\dagger \sigma^3 \mu^3\left[\tau^1(\partial_1+i A_1)-\tau^2(\partial_2+i A_2)\right]\Psi$~\cite{hermeleAlgebraicSpinLiquid2005}.
This type of perturbation is in fact expected to be irrelevant (e.g. in the large-$N_f$ expansion)~\cite{hermeleAlgebraicSpinLiquid2005}.
It also does not impact the primary effect of the combined term $H_z^{(1)}+H_z^{(2)}$, which is to open a gap for fermion excitations [Fig.~\ref{fig:SI_glidermass}(h)] and consequentially confine the $U(1)$ gauge field.

\subsection{Kagome Brillouin zones}
For completeness, we recount some details of the kagome geometry that can occasionally cause confusion.
The kagome lattice is composed of three interlacing triangular sublattices, which we label $\mathfrak{a}$, $\mathfrak{b}$, and $\mathfrak{c}$.
We take units where the spacing between nearest-neighbors is $a=1$, then choose the positions within the unit cell to be $\mb{\rho}_{\mathfrak{a}} = (0,0)$, $\mb{\rho}_{\mathfrak{b}}=(\sqrt{3}/2, 1/2)$, and $\mb{\rho}_{\mathfrak{c}} = (0,1)$, and the basis vectors as $\mb{a}_1 = (\sqrt{3}, 1)$, $\mb{a}_2 = (0,2)$.
With this choice, $\mb{a}_2$ is along the periodic direction of the YC-oriented cylinders (e.g. YC12 in main text Fig. 1(b)). 
Each site of the lattice is then at the position $\mathbf{r}_i = m_i \mb{a}_2 + n_i \mb{a}_2 +\rho_i$, where $(m_i,n_i)\in\mathbb{Z}^2$.
The corresponding reciprocal vectors are  $\mb{b}_1 = (\pi, 0)$ and $\mb{b}_2 =(2\pi/\sqrt{3}, -\pi/\sqrt{3})$, so that $\mb{a}_i \cdot \mb{b}_j  = 2\pi \delta_{ij}$. 
These span a triangular Bravais lattice, and momentum space can hence be divided into hexagonal Brillouin zones (BZ), shown as the black hexagon in Fig.~\ref{fig:SI_spinonschematic}.
The high-symmetry points of this \textit{inner}  BZ are $\Gamma=(0,0)$,  $M_I^{(i)} = \pm \pi \cdot (\cos(2\pi i/6),\,\sin(2\pi i/6))$, where $i=1,2,3$; and $K_I^\pm = \pm (\pi, \pi/\sqrt{3})$. 
It is often useful in studies of kagome systems to also consider the \textit{extended} BZ, shown as the light gray hexagon in Fig.~\ref{fig:SI_VBS_DSL}.
This corresponds to the BZ of the finer triangular lattice of which the kagome lattice is a subset (i.e. the one which arises by adding a site to the center of each hexagon).
The extended BZ has twice the linear dimension of the inner BZ, and we denote its high symmetry points as $M_E^{(i)} = 2 M_I^{(i)}$ and $K_E^\pm = 2 K_I^\pm$.
A typical appearance of the extended BZ is in the sublattice-unresolved, equal-time, spin structure factor,
\begin{equation}
S^{\alpha\beta}(\mb{k}) =  \frac{1}{N}\sum_{i,j} e^{i \mb{k}\cdot(\mb{r}_i-\mb{r}_j)}  \langle S^\alpha_i S^\beta_j \rangle,
\end{equation}
which is, for instance, probed in neutron scattering experiments.
The fermion billinear excitations of the $U(1)$ DSL have momentum at the $\Gamma$ or $M_I$ points.
They are thus most cleanly probed in the \textit{inner} BZ; the Friedel oscillations discussed in the main text are an example of this.
Another example is the sublattice-resolved dynamical spin structure factor,
\begin{equation}
\mathcal{S}_{\lambda\mu}^{\alpha\beta}(\mb{k}, \omega) = \frac{1}{2\pi}\sum_{\mathbf{R}} \int_{-\infty}^{\infty}e^{i(\omega t -\mb{k}\cdot\mb{r})}\langle\sigma_{\mb{R},\lambda}^\alpha(t)\sigma_{\mb{0},\mu}^\beta(0)\rangle, 
\end{equation}
where $\lambda,\mu$ are sublattice indices, for which we plot the mean-field spinon contribution in Fig.~\ref{fig:SI_spinonschematic}(d).
The gapless weight at the $\Gamma$ and $M_I$ points respectively correspond to intra- or inter-node particle-hole scattering processes.


\section{Edge states and bulk modes on infinite strips}

In the main text, we discussed how certain terminations of the kagome lattice hosts specific low-energy modes, as a consequence of the presence of Dirac-cones in the mean-field spinon spectrum. 
In this section, we elaborate on the origin of these low-energy modes, and show that they can be derived from a continuum low-energy theory that describes the Dirac cones, together with appropriate boundary conditions that describe the edge terminations.
Alternatively, one may solve the tight-binding Hamiltonian in Eq.~\eqref{eq:HmfPiFlux} on an infinite strip to obtain these low-energy modes, as shown in Fig. 4 in the main text. 
Our analytic description, based on the Dirac spectrum, can precisely capture not only the energetics of such modes, but also whether they are localized or not --- indicating a strong connection of these modes with the Dirac spectrum of spinons.

Our strategy is to first derive an effective continuum low energy model for the bands near the Dirac cones. 
We then solve the resulting Dirac equation for the spectrum using appropriate boundary conditions, that depends on the specific termination of interest. 
For convenience, we rotate the unit cell shown in FIG. \ref{fig:SI_spinonschematic}.
Note that the unit cell now has 6 sites, because of the presence of $\pi$ flux through each hexagonal plaquette.
After taking a Fourier transform of the tight-binding Hamiltonian in Eq.~\eqref{eq:HmfPiFlux}, the two Dirac points lie at $\mb{Q}=\pm(0,\frac{\pi}{2})$. 

We begin by expanding the Hamiltonian near $\mb{Q}$ to linear order
in momentum (in the spirit of a $\mb{k} \cdot \mb{p}$ expansion).
Defining $D_+(\mb{k}) = H(\mb{Q}+\mb{k})-H(\mb{Q})$, we find
\begin{equation}
D_{+}(\mb{k}) = \begin{pmatrix}
0 & 0 & -(k_x+\sqrt{3}k_y) & 0 & 4ik_x  & 0\\
0 & 0 & 0 & 0 & 0 & -(k_x+\sqrt{3}k_y)\\
-(k_x+\sqrt{3}k_y) & 0 & 0 & 0 & 3k_x-\sqrt{3}k_y & 0\\
0 & 0 & 0 & 0 & 0 & (k_x+\sqrt{3}k_y)\\
-4ik_x & 0 & 3k_x-\sqrt{3}k_y & 0 & 0 & 0\\
0 & -(k_x+\sqrt{3}k_y) & 0 & (k_x+\sqrt{3}k_y) & 0 & 0
\end{pmatrix}
\end{equation}
We can find a similar expression for $D_-(\mb{k})$ by expanding around $\mb{Q}_-$. 
At each Dirac point there are two dgenerate bands, so the low-energy Hilbert space manifold is four dimensional and is spanned by the following vectors: 
\begin{align}
v_{1}^{+} &= \frac{1}{\sqrt{6}}\left(e^{-2\pi i/3} ~\sqrt{2}e^{-\pi i/12} ~i ~e^{\pi i/3} ~0 ~1\right)^{T}\\
v_{2}^{+} &= \frac{1}{\sqrt{6}}\left(e^{11\pi i/12} ~0 ~e^{-3\pi i/4} ~e^{11\pi i/12} ~\sqrt{2}e^{\pi i/3} ~e^{-\pi i/4}\right)^{T} \\
v_{1}^{-} &= \frac{1}{\sqrt{6}}\left(e^{2\pi i/3} ~\sqrt{2}e^{\pi i/12} ~-i ~e^{-\pi i/3} ~0 ~1\right)^{T} \\
v_{2}^{-} &= \frac{1}{\sqrt{6}}\left(e^{-11\pi i/12} ~0 ~e^{3\pi i/4} ~e^{-11\pi i/12} ~\sqrt{2}e^{-\pi i/3} ~e^{\pi i/4}\right)^{T}
\end{align}
Each vector has six components for the six atoms in the unit cell $(A,B,C,D,E,F)$. The effective Hamiltonian's matrix elements (in first order perturbation theory) are $H_{ij} = \bra{v_i}D(\mb{k})\ket{v_j}$. 
This leads to the following effective Hamiltonian description of the low-energy theory:  
\begin{equation}
H(\mb{q}) = \sqrt{2}\left(q_{x}\sigma_{y}+q_{y}\tau_{x}\sigma_{x}\right),
\end{equation}
where $\sigma$ is a band index and $\tau$ is the valley index. We will use this effective Hamiltonian to study systems that are finite in one direction and infinite in the other to describe the resulting bulk and edge modes.

We consider the systems with hexagon and flat terminations (as defined in the main text) to illustrate the utility of our model. 
The system with the hexagon termination shows chiral zero-energy modes that are extended throughout the lattice while the flat termination shows exponentially localized edge states. 
Both these features can be understood from stuyding the continuum Hamiltonian near the Dirac cones. 

Starting with the hexagon termination, we take the strip to be infinite in the $x$ direction and finite in the $y$ direction. This implies that we can substitute for $q_y$ as $q_y = -i\partial_y$ while $q_x$ remains a good quantum number. This leads to the following differential equations in the $\tau = +$ valley,
\begin{align}
i\sqrt{2}\left(-q_x + \partial_{y}\right)\phi_2^{+}(y) = \varepsilon \phi_1^{+}(y) &\\
i\sqrt{2}\left(q_x + \partial_{y}\right)\phi_1^{+}(y) = \varepsilon \phi_2^{+}(y)
\end{align} 
which leads to the decoupled equation
\begin{equation}
\left(q_x^2-\partial_y^2\right)\phi_1^{+}(y) = \frac{\varepsilon^2}{2}\phi_1^{+}(y)
\end{equation}
The general solution to this equation is,
\begin{equation}
\phi_1^{+}(y) = Ae^{izy}+Be^{-izy},
\end{equation}
$z^2 = \frac{\varepsilon^2}{2}-q_x^2$ which gives the energy as a function of $q_x$ to be $\varepsilon = \sqrt{2(q_x^2+z^2)}$.
The role of the boundary conditions is to then determine what values of $z$ are allowed.

The appropriate boundary conditions are determined by setting the wave function to zero at the lattice sites that are extensions of the atoms at the ends of the strip. In the low energy theory, the wave function at a band $\mu$ is given by
\begin{equation}
\Psi_{\mu}(\bm{r}) = e^{i\bm{Q}\cdot\bm{r}}\psi_{\mu}^{+}(\bm{r}) + e^{-i\bm{Q}\cdot\bm{r}}\psi_{\mu}^{-}(\bm{r})
\end{equation}
Since the system is infinite in the $x$ direction, we can write the wave function $\psi(x,y)$ as $\psi(x,y) = e^{iq_{x}x}\phi(y)$. 
The ends of the strips are found at $y = 0$ and $y = W$, where $W$ is the width of the strip.
At $y = 0$, we get 
\begin{equation}
0 = e^{iq_x x}(\phi_1^{+}(0)+\phi_1^{-}(0))
\end{equation}
To satisfy the above equation for all $x$, we must have $\phi_1^{+}(0)+\phi_1^{-}(0) = 0$.

At $y = W$, we get \begin{equation}
0 = e^{iq_x x}\left(e^{i\frac{\pi}{2}W}\phi_1^{+}(W)+e^{-i\frac{\pi}{2}W}\phi_1^{-}(W)\right)
\end{equation}

Let us consider propagating modes in the $y$-direction, such that $\phi_1^{+}(y) = e^{izy}$, $\phi_1^{-}(y) = -e^{-izy}$, satisfying the boundary condition $\phi_1^{+}(0)+\phi_1^{-}(0)= 0$.
In order to satisfy the boundary conditions at $y = W$ we must have 
\begin{equation}
e^{izW} = e^{-i\pi W}e^{-izW}
\end{equation}

which we can solve for $z$ to get 
\begin{equation}
z = -\frac{\pi}{2} + \frac{n\pi}{W}~~~~\text{for}~n\in \mathbb{Z} 
\end{equation}
Now, recall the energy on the finite strip is given by $\varepsilon = \sqrt{2(q_x^2+z^2)}$, where $q_x = 0$ corresponds to the Dirac point. 
So, the spectrum can be gapless for $n = \frac{W}{2}$. 
However, since $n \in \mathbb{Z}$, such a solution is only possible if $W$ is even, which corresponds to the hexagon termination. 
Additionally, note that such a mode is delocalized all over the bulk, consistent with numerical observations.

We now turn to the flat termination, which is obtained by having the periodic direction along $y$ and the strip begins at $x = 0$ and ends at $x = W$. 
In this termination, the wave function vanishes at the sublattice E at $x = 0$ and at sublattice B at $x = W$. 
Writing the wavefunction as $\psi(x,y) = e^{i k_y y} \phi(x)$, the vanishing of the required components of $\psi(x,y)$ at $x= 0$ and $x = W$ translates to the boundary conditions 
\begin{equation}
\phi_1^{+}(x = 0) = \phi_1^{-}(x = 0) = \phi_2^{+}(x = W) = \phi_2^{-}(x = W) = 0.
\end{equation}
We note that this is identical to the boundary condition of the zigzag termination in graphene. We seek to find solutions that are decaying in the bulk so they take the form $\phi_1^{+}(x) = Ae^{zx}+Be^{-zx}$, where $z^2 = \frac{\varepsilon^2}{2}-q_y^2$. To satisfy these boundary conditions, we find that we need to solve the transcendental equation,
\begin{equation}
\label{eq:transcdental equation}
\frac{k_y - z}{k_y + z} = e^{-2Wz}
\end{equation}
Finding real solutions for Eq.~\eqref{eq:transcdental equation} amounts to finding states that are localized on the edge (edge states). 
Real solutions are found for $k_y > k_c = 1/W$ with no real solutions for $k_y < 0$. 
These analytical results agree with the tight-binding numerics on a strip, where surface states are only found to the right/left of the Dirac points, and are absent on the other side.
\section{Additional iDMRG results}\label{app:kagome}

\subsection{iDMRG methodology}\label{app:idmrg}
Here we comment on the application of iDMRG to the $H_{\rm{dXY}}$ problem.
There are three key approximations that must be made.
First, the true two-dimensional lattice must be compactified to a quasi-one-dimensional system, by choosing periodic boundary conditions along some vector $\vec{r}_{\rm mod}$ .
This corresponds to an infinitely long cylinder with  finite circumference, $W = \norm{\vec{r}_{\rm mod}}$, which is typically six to twelve sites.
Second, the power-law interactions of $H_{\rm{dXY}}$ cannot be exactly represented as a matrix product operator (MPO), so we introduce an additional length scale, $R_{\rm{int}}$, for the interaction range.
There are various schemes to perform this MPO compression, but in practice we found that the ground state is typically insensitive to the longest-range interactions, and so we simply set all couplings to zero beyond some maximum radius $R_{\rm{int}}\lesssim W/2$.
Finally, the matrix product state (MPS) representation is necessarily restricted to a finite bond dimension $d$. 
The true ground state of the system corresponds to the limit in which $W$, $R_{\rm int}$, and $d$ are all three taken to infinity.
While powerful, iDMRG is not an infallible technique for studying two-dimensional systems.
This is especially so for frustrated magnetic systems, with many competing orders and the possibility of strong finite size effects.
To help ensure the reliability of the iDMRG results, we proceed as follows.

First, we work with a broad set of cylindrical geometries, which correspond to different choices of the periodic vector. 
This helps to check against the emergence of ordered states (e.g. $\sqrt{3}\times\sqrt{3}$) that are energetically frustrated by certain unfavorable geometries. 
However, we are generally speaking unable to rule out ordered states with unit cells larger than the circumference of our cylinders (as may be the case for certain large-scale valence bond solids). 
The largest cylinder we study is the YC12 geometry shown in the main text.
In somewhat different models, it is known that long-range antiferromagnetic interactions have a tendency to stabilize mesoscopic ordering patterns; assessing this possibility appears beyond the capabilities of iDMRG or any numerically-exact technique we know of, but may be an interesting topic for future study with variational methods. 

Second, we attempt to safeguard against iDMRG's tendency to get stuck in local energy minima. 
We believe this is especially important for $H_{\rm{dXY}}$, which has many competing interactions and proximate phases.
To do so, on each cylinder we first initialize up to 40 random states.
These states are either random singlet coverings of the unit cell, or the result of a moderate depth random unitary circuit projected to $\chi=512$. 
We use real wavefunctions for these; separately, we also use time-reversal-breaking (Kalmeyer-Laughlin) chiral spin liquid wavefunctions as candidate initial states.
We perform a first run of iDMRG on these states up to an intermediate bond dimension of $\chi=2048-4096$.  
Then we sort them into equivalency classes based off their energy and the structure of their spin- and momentum-resolved entanglement spectrum. 

We then run iDMRG a second time on single representatives of these classes, scaling to much larger bond dimension, $\chi=6144-10240$.
In our iDMRG sweeps, we generally increment the bond dimension in units of $1024$ or $2048$, and make use of a mixer (for two-site iDMRG) or subspace expansion (for single-site iDMRG) whenever the bond dimension is increased. 
Occasionally there are many states competing at very small energy scales, with relative ordering that can change with increasing bond dimension. 
Nevertheless, in most cases we are able to identify a unique lowest energy state. 
We focus on this lowest state for each cylinder.

\subsection{Connection to Heisenberg model}

\begin{figure}
	\includegraphics{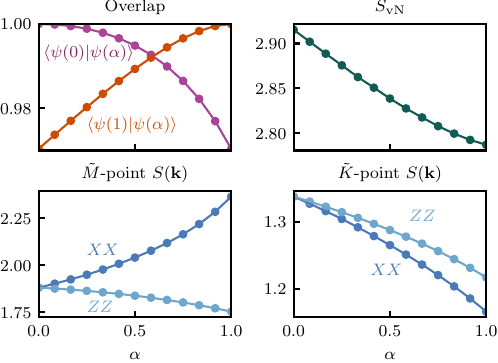}
	\caption{\label{fig:SI_heis}   Interpolation from  $H_{\rm nn}$ ($\alpha=0$) to $H_{\rm dXY}$ ($\alpha=1$) on the infinite YC8 cylinder (bond dimension $d=3072$).  (a) The ground state overlap (per three-site kagome unit cell) remains large, with no indication of any sharp dip that would be seen at a phase transition. (b) Going from $H_{\rm{nn}}$ to $H_{\rm{dXY}}$, the entanglement entropy smoothly decreases by a small amount. (c)  The largest change is at the $\tilde{M}$ point, for which the $XX$ correlations are enhanced, and the $ZZ$ correlations diminish. (d) At the $\tilde{K}$ point, correlations are slightly reduced for both $XX$ and $ZZ$. 
	}
\end{figure}

%
We consistently find spin liquid states that are very similar to that of the antiferromagnetic Heisenberg model, $H_{\mathrm{nn}} = (J/2)\sum_{\langle i,j\rangle} \vec{\sigma_i}\cdot\vec{\sigma_j}$,
with the sum taken over just nearest-neighbors.
To explicitly test this relation, we consider the continuous interpolation $\alpha H_{\rm nn} + (1-\alpha)H_{\rm dXY}$, and compute the iDMRG ground state as $\alpha$ varies from 0 to 1, using the YC8 cylinder.
We find a smooth crossover between the two endpoints, with no sign of any intervening phase transition (Fig. ~\ref{fig:SI_heis}).
Thus, the quantum spin liquid we find is likely in the same phase as the extensively characterized $H_{\rm nn}$ ground state.

\subsection{Global XY phase diagrams}\label{sec:SI_global}

The dipolar XY Hamiltonian, $H_{\rm dXY}$, can be located within a broader phase diagram of generic antiferromagnetic $XY$ models on the kagome lattice.
To this end, we consider the three-coupling model,
\begin{equation}
H_{3J} = J_1 \sum_{\langle i,j \rangle} (X_i X_j + Y_i Y_j) 
+ J_2 \sum_{\langle \langle i,j \rangle \rangle} (X_i X_j + Y_i Y_j) 
+  J_3 \sum_{\langle \langle \langle i,j \rangle \rangle \rangle} (X_i X_j + Y_i Y_j) ,
\end{equation}
where the three sums are taken over nearest-neighbors, next-nearest-neighbors, and next-next-nearest neighbors, respectively.
Note that the $J_3$ terms have two geometric types on the kagome lattice, across hexagons and across bowties, which we include both of.
%

\begin{figure}
	\includegraphics{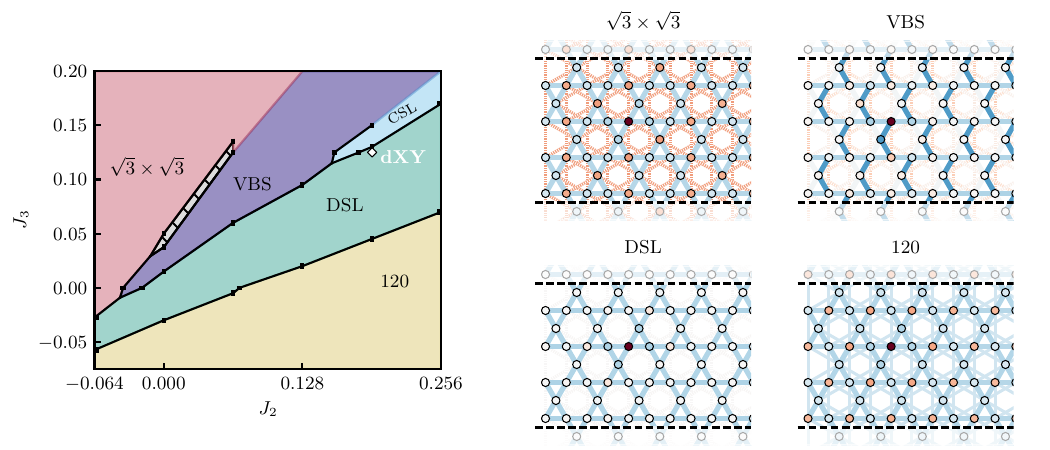}
	\caption{\label{fig:SI_global3} \textbf{Left:} Global phase diagram of the $J_1-J_2-J_3$  model on the kagome XC8 infinite cylinder (bond dimension $d=3072$). White diamond marks the $J\propto1/r^3$ point. Black markers are phase transitions determined from vertical or horizontal cuts; black lines interpolate between them. Gray hatched area is a thin critical region where the VBS order vanishes and the $\sqrt{3}\times\sqrt{3}$ correlations are short-ranged. \textbf{Right:} Real-space correlation patterns of the various phases. Bonds show nearest- and next-nearest-neighbor $\langle \sigma^x_i \sigma^z_j \rangle$ correlations; sites show correlations $\langle \sigma^x_i \sigma^x_0\rangle$ with a fixed site (dark red circle).}
\end{figure}

%
With only nearest-neighbor interactions, the ground state of the XY kagome AFM is likely to be a spin liquid. 
In fact, numerical studies have generally found that the nearest-neighbor kagome XXZ model, $H_{\rm XXZ} = - \frac{J}{2} \sum_{\langle i j \rangle}  X_i X_j + Y_i Y_j + \Delta Z_i Z_j$, 
has a constant spin liquid ground state for any $\Delta > -1/2$~\footnote{Note our convention for $J$ here, which is chosen to match $H_{\rm dXY}$, but differs from the standard one in the literature.}.
The XY  point ($\Delta=0$), 
\begin{equation}
H_{\rm nnXY} = -\frac{J}{2}  \sum_{\langle i j \rangle}  X_i X_j + Y_i Y_j
\end{equation}
has been specifically studied on occasion~\cite{lauchliQuantumSimulationsMade2015, huVariationalMonteCarlo2015}.
For example, in Ref.~\cite{lauchliQuantumSimulationsMade2015}, L\"auchli and Moessner showed via exact diagonalization that the low-lying spectra of the XY and the $\Delta=1$ SU(2) kagome models are essentially identical up to rescaling.  
This suggests that the $H_{\rm nnXY}$ ground state is also in the $U(1)$ DSL phase, which we confirmed through iDMRG flux insertion calculations.
Taking a series of cuts through the $J_2, J_3$ phase space, and setting $J_1=1$, we find the iDMRG ground state of $H_{3J}$ on the infinite XC8 cylinder.
The results of this calculation are summarized in Fig.\;\ref{fig:SI_global3}. 
The addition of $J_2$ favors the formation of the 120-ordering state, which spontaneously breaks the $U(1)$ symmetry but is translation invariant. 
At large $J_3$, one instead finds the $\sqrt{3}\times \sqrt{3}$ state, which again has XY LRO but also enlarges the real-space unit cell to be 9 sites.
When $J_2$ and $J_3$ are of comparable magnitude, competition  between the two gives way to either a spin liquid state, or to a valence bond solid (VBS) characterized by spatially non-uniform, short-range correlations. 
The interactions of the dipolar XY model correspond to the point $J_2/J_1 = 1/(3\sqrt{3}) \approx 0.192$ and $J_3/J_1 = 1/8$, within this region of competition.
In fact, it is very close to a second-order phase boundary between \textit{two} quantum spin liquids.
The first of these is the presumed-DSL, which is smoothly connected to the nearest-neighbor point.
The second is a Kalmeyer-Laughlin chiral spin liquid (CSL) state that spontaneously breaks time-reversal-symmetry, and has chiral semion topological order. 
On this cylinder, the dipolar point appears to be slightly on the DSL side; larger cylinders and the inclusion of longer-range interactions further stabilize the DSL state.

\subsubsection{Five-coupling model}

We now zoom in on the parameter space surrounding $H_{\rm{dXY}}$. 
In particular, we fix $(J_2,J_3)=(0.19, 0.125)$ to their $1/r^3$ values, and now perturb the system with couplings $(J_4, J_5)$, coupling sites at distances $(r_4,r_5)=(\sqrt{7},3)$.
Near the dipolar values $(J_4, J_5) = (0.054, 0.037)$, there is an enlarged region of the DSL phase. 
Thus, although the influence of $(J_2, J_3)$ bring the system close to the CSL phase, the longer-range antiferromagnetic interactions provide sufficient frustration to return towards the original DSL.

\begin{figure}
	\includegraphics{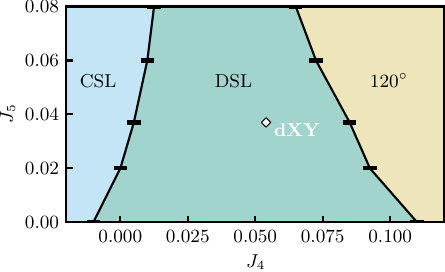}
	\caption{\label{fig:SI_global5} Phase diagram of the five-coupling  model on the XC8 kagome cylinder, fixing $(J_2, J_3) = (0.192, 0.125)$ to match the dipolar Hamiltonian. White diamond marks the dipolar point.
	}
\end{figure}

\subsubsection{Longer-range interactions}
In practice we checked that adding the additional unambiguous couplings for XC8 ($J_6=0.024$, $J_7=0.021$) did not lead to any considerable impact on the ground state, i.e. the long-range interactions appear to have stabilized the system within the DSL phase.
For the YC12 and YC8-2 calculations discussed in the main text, we used an $R_{max}= 3.7$ cutoff, which corresponds to the inclusion of seven different couplings.
The first excluded coupling in this case is $J_8 = 1/64$, between sites separated by distance 4. 
We chose this as a reasonable balance between long-range coupling effects (which are eventually irrelevant, i.e. in the renormalization group sense [Sec.~\ref{app:dsl}]) and computational cost. 
We also believe that finite-$W$ effects are dominant over the impact of the excluded interactions~\cite{ferrariGaplessSpinLiquids2021}. 
%

\section{Additional finite-size DMRG}\label{app:finite}
\subsection{The \texorpdfstring{$N=80$}{80} cluster: VBS vs DSL}

\begin{figure}[b]
	\includegraphics{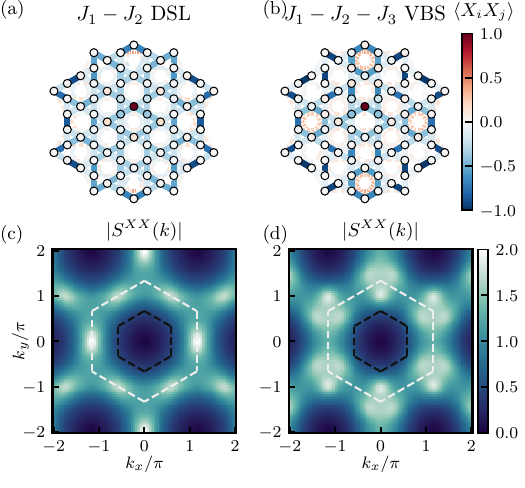}
	\caption{\label{fig:SI_VBS_DSL} Finite DMRG results for an $N=80$ cluster, at bond dimension $d=4096$. (a,b) Real-space $\langle\sigma^x_i \sigma^x_j\rangle$ correlations  in the (a) DSL and (b) VBS phases of the $J_1-J_2-J_3$ model. (c,d) The respective Fourier transforms reveal the $M_E$ weight is strong for the DSL but absent in the VBS, and so can distinguish the two.  
	}
\end{figure}

In the thermodynamic limit, a VBS phase is most simply distinguished from the DSL by local correlations which spontaneously break spatial symmetries.
This is subtle to judge on finite, open-boundary clusters, so we seek an alternative.
This turns out to be straightforward, at least for the particular VBS seen in the global XY phase diagram.
In Fig.~\ref{fig:SI_VBS_DSL}, we show $\langle \sigma^x_i \sigma^x_j\rangle$ and its Fourier transforms for the $J_2=0.025$ DSL and the $(J_2, J_3) = (0.128, 0.125)$ VBS on an $N=80$ site cluster: the DSL correlations are predominantly at the $M_E$ points of the extended B.Z., while for the VBS the weight is concentrated near the corners.
This distinction can be further amplified by omitting the nearest-neighbor correlations (not shown).

\subsection{Interstitial defect for other phases}

\begin{figure*}
	\includegraphics{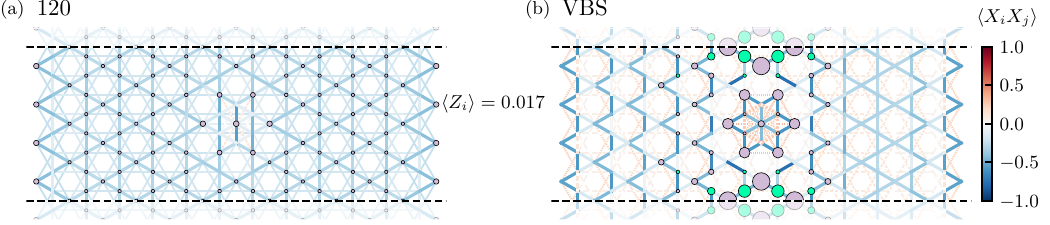}
	\caption{\label{fig:SI_interstitial}  The effect of an interstitial site on other kagome XY phases, as seen in finite-cylinder DMRG (bond dimension $d=3072$). Bonds again indicate $\sigma^x\sigma^x$ correlations, while the circles show single-body $\langle\sigma^z_i\rangle$ (lavender for positive, green for negative, with area proportional to $|\langle \sigma^z_i\rangle|$).
	}
\end{figure*}

In the main text, we proposed using an interstitial defect atom as a way to introduce an additional spin-1/2 into the system.
To our knowledge, such a defect has not been previously studied in kagome numerics.
In Fig.~\ref{fig:SI_interstitial}, we show finite-cylinder DMRG results for the effect of this defect on other phases of the  $J_1-J_2-J_3$ model.
In the $120^\circ$ state [Fig.~\ref{fig:SI_interstitial}(a)], the excess spin-1/2 mostly spreads uniformly over the system, with some inhomogeneity near the inserted site.
Interestingly, this acts as a defect in the $q=0$ ordering, which may lead to more complex effects on open clusters.
The VBS [Fig.~\ref{fig:SI_interstitial}(b)] behaves somewhat similarly to the DSL: the twelve spins near the interstitial site become very strongly correlated with one another, and largely decouple from the rest of the system.
There are large oscillations in $\sigma^z_i$ near the defect; the particular eight-spin ``diamond'' resonance is possibly an artifact of the cylinder geometry.

\section{Note on adiabaticity}
The staggered $\sigma^z$ field ramp is an example of the ``end-critical protocol'' discussed in Ref.~\cite{chandranKibbleZurekProblemUniversality2012}.
As noted in the main text, with sufficiently slow power-law-type ramps $\delta(t) \sim t^{-p}$, the $\delta=0$ critical point ($U(1)$ DSL) can be approached arbitrary closely without losing adiabaticity.
Here we give a short analysis of the exponential ramp used in the TDVP simulation.
The Ref.~\cite{chandranKibbleZurekProblemUniversality2012} criteria for adiabaticity is that the change in correlation time $\Delta \xi_t$ over a period of correlation time $\xi_t$ is much smaller than the correlation time $\xi_t$ itself. 
In other words, we need $\Delta \xi_t = \dot{\xi}_t \xi_t \ll \xi_t$, or $\dot{\xi}_t \ll 1$. 
For an exponential ramp of a relevant parameter $\delta = \delta_0 e^{-t/\tau}$, the instantaneous correlation time is given by 
\begin{equation}
\xi_t \sim [\delta(t)]^{-\nu z} = \delta_0^{-\nu z} e^{\nu z t/\tau} \implies \dot{\xi}_t =  \frac{\nu z}{\tau} \delta_0^{-\nu z} e^{\nu z t/\tau}
\end{equation}
If we want to estimate the time $t_Q$ such that $ \dot{\xi}_{t=t_Q} = 1$, we find that
\begin{equation}
\dot{\xi}_{t = t_Q} = 1 =  \frac{\nu z}{\tau} \delta_0^{-\nu z} e^{\nu z t_Q/\tau} \implies t_Q = \frac{\tau}{\nu z} \left( \ln(\tau/\nu z) + \nu z \ln(\delta_0)\right)
\end{equation}
Unlike the power law case discussed in Ref.~\cite{chandranKibbleZurekProblemUniversality2012}, in this case there is a finite time-scale $t_Q$ after which the system falls out of equilibrium.  
However, for large $\tau$ this time-scale diverges as $t_Q \sim \tau \ln(\tau)$ so we can ramp down to very small values of $\delta$ over a time-scale $\sim A\tau$ where $A < \ln(\tau)$ and still be approximately adiabatic. 

\section{Dipolar XY antiferromagnets}\label{app:dxy_afm}

\begin{figure}
	\includegraphics{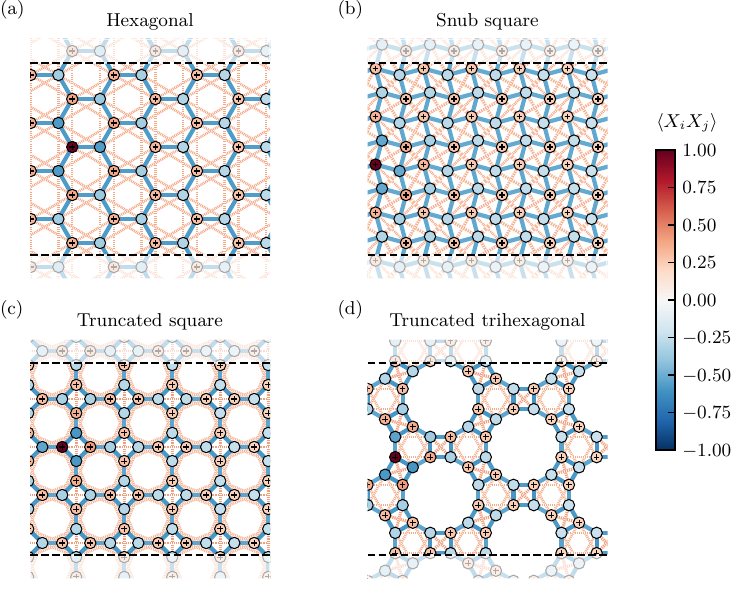}
	\caption{\label{fig:SI_collinear} Enlarged depiction of the various collinear ordering patterns supported by $H_{\rm{dXY}}$, as found in iDMRG ($d=2048$).  Bonds show ground state correlations between nearest-neighbor and next-nearest neighbors sites; red dashed lines correspond to ferromagnetic $\langle \sigma^x_i \sigma^x_j\rangle>0$, while blue solid lines are antiferromagnetic.  Circles show $\langle \sigma^x_i \sigma^x_0\rangle$ correlations with a fixed site (dark red circle), and those with $\langle \sigma^x_i \sigma^x_0 \rangle >0$ are also marked with a '$+$' sign.  
	}
\end{figure}

\begin{figure}
	\includegraphics{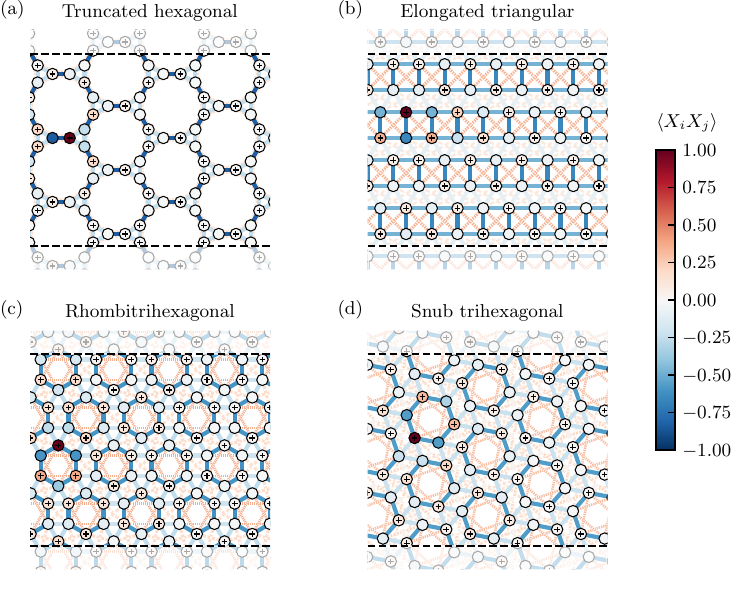}
	\caption{\label{fig:SI_localsinglets} Enlarged depiction of the various local singlet states (plot uses same scheme as in Fig.~\ref{fig:SI_collinear}). In the condensed matter literature, these lattices occasionally go under different names (``Fisher'' for (a), ``ruby'' for (c), and ``maple leaf'' for (d)).
	}
\end{figure}

Here we give some additional information on the non-kagome-lattice ground states of the $H_{\rm{dXY}}$ antiferromagnet.  
(For ferromagnetic interactions, $H_{\rm{dXY}}$ is generally expected to host a simple, lattice-independent $U(1)$ symmetry breaking ground state---notably, one stable to finite temperature due to the long-range interactions~\cite{chenContinuousSymmetryBreaking2023}.)
The Archimedean lattices~\cite{grunbaumTilingsPatterns2016} with collinear ordering are shown in  Fig.~\ref{fig:SI_collinear}.
The square lattice is also in this class~\cite{chenContinuousSymmetryBreaking2023}.
The basic signature of spontaneous $U(1)$ symmetry breaking in our iDMRG calculations (which use $U(1)$-symmetric tensors) is off-diagonal long-range order in the $\sigma^x$ (equivalently, $\sigma^y$)  correlations.
In the depicted geometries, the ground state $\langle \sigma^x_i \sigma^x_j \rangle$ correlations are large and roughly constant out to long distances, before slowly decaying due to the quasi-1d infinite cylinder geometry and the finite MPS bond dimension.
The correlations exhibit an obvious sign structure characteristic of a simple, bipartite Ne\'el antiferromagnet. 
Such order is frustrated at the nearest-neighbor level for the snub square tiling, but only at longer distances (and hence weaker coupling) in the other cases.

Local geometric frustration penalizes collinear ordering.
For some lattices with complex unit cells, no ordering occurs at all, and the ground state of $H_{\rm dXY}$ is a short-range-entangled paramagnet (Fig.~\ref{fig:SI_localsinglets}).
This type of state is similar to a valence bond solid, but strictly speaking is not a true ``solid'' as no spatial symmetries are spontaneously broken.
Such a fully symmetric paramagnetic state is possible because the unit cells contain an even number of spins.
Note that although our Hamiltonian is only $U(1)$ symmetric, the ground state of an antiferromagnet XY interaction between two spin-1/2 is identical to an $SU(2)$ singlet; this relation no longer holds exactly for our many-body wavefunctions, but we do find that the local correlations in this class of states are largely spin-isotropic (not shown).

\newpage
\bibliography{kgm_paper}